\documentclass[fleqn,usenatbib]{mnras}

\usepackage{newtxtext,newtxmath}
 
\usepackage[T1]{fontenc}
\usepackage{ae,aecompl}

\usepackage{graphicx}	% Including figure files
\usepackage{amsmath}	% Advanced maths commands
\usepackage{amssymb}	% Extra maths symbols

\newcommand{\hdtwenty}{HD~209458~b}
\newcommand{\hdeighteen}{HD~189733~b}

\newcommand{\Teff}{T_{\textrm{eff}}}

\newcommand*\chem[1]{\ensuremath{\mathrm{#1}}}
\newcommand{\kzz}{K_{zz}}

\newcommand{\degrees}{^\circ}
\newcommand\psj{Planet. Sci.~J.}

\title[The role of photochemistry]{Grid of pseudo-2D chemistry models for tidally locked exoplanets -- II. The role of photochemistry}

\author[R. Baeyens et al.]{%
Robin Baeyens,$^{1}$\thanks{E-mail: robin.baeyens@kuleuven.be}
Thomas Konings,$^{1}$
Olivia Venot,$^{2}$
Ludmila Carone$^{3, 4}$ and\newauthor%
Leen Decin$^{1}$
\\
$^{1}$Institute of Astronomy, KU Leuven, Celestijnenlaan 200D, 3001 Leuven, Belgium\\
$^{2}$Universit\'e de Paris and Universit\'e Paris-Est Cr\'eteil, CNRS, LISA, F-75013 Paris, France\\
$^{3}$Max-Planck-Institut f\"ur Astronomie, K\"onigstuhl 17, 69117 Heidelberg, Germany\\
$^{4}$Centre for Exoplanet Science, School of Physics \& Astronomy, University of St Andrews, North Haugh, St Andrews, KY169SS, UK
}

\date{Accepted XXX. Received YYY; in original form ZZZ}

\pubyear{2022}

\begin{document}
\label{firstpage}
\pagerange{\pageref{firstpage}--\pageref{lastpage}}
\maketitle

\begin{abstract}
Photochemistry is expected to change the chemical composition of the upper atmospheres of irradiated exoplanets through the dissociation of species, such as methane and ammonia, and the association of others, such as hydrogen cyanide. Although primarily the high altitude day side should be affected by photochemistry, it is still unclear how dynamical processes transport photochemical species throughout the atmosphere, and how these chemical disequilibrium effects scale with different parameters. In this work we investigate the influence of photochemistry in a two-dimensional context, by synthesizing a grid of photochemical models across a large range of temperatures. We find that photochemistry can strongly change the atmospheric composition, even up to depths of several bar in cool exoplanets. We further identify a sweet spot for the photochemical production of hydrogen cyanide and acetylene, two important haze precursors, between effective temperatures of 800 and 1400~K. The night sides of most cool planets ($\Teff < 1800$~K) are shown to host photochemistry products, transported from the day side by horizontal advection. Synthetic transmission spectra are only marginally affected by photochemistry, but we suggest that observational studies probing higher altitudes, such as high-resolution spectroscopy, take photochemistry into account.
\end{abstract}


%%%%%%%%%%%%%%%%%%%%%%%%%%%%%%%%%%%%%%%%%%%%%%%%%%

%%%%%%%%%%%%%%%%% BODY OF PAPER %%%%%%%%%%%%%%%%%%

% See \texttt{mnras\_sample.tex} for a more complex example, and \texttt{mnras\_guide.tex} for a full user guide.

%%%

\section{Introduction}

% what is photochemistry
The day-side hemispheres of tidally locked exoplanets are constantly being irradiated by ultraviolet (UV) radiation originating from the stellar host. This irradiation in turn leads to photochemistry in the planet's atmosphere, by which we mean the set of chemical reactions and reaction mechanisms that incorporate interactions with a photon. In the Solar System planets, photochemistry is ubiquitous \citep[e.g.][and references therein]{Yung1999, Strobel2005, Moses2020}. Because of their strong one-sided irradiation, gaseous tidally locked exoplanets are likewise expected to have their chemical composition, and their potential transmission signatures, altered by photochemistry. Indeed, effects of disequilibrium chemistry are just starting to be probed through exoplanet observations \citep{Baxter2021, Roudier2021}.

% Spefically on photochemical hazes
Another potential impact of photochemistry is the formation of hazes. Given the common detections of haze in exoplanetary atmospheres \citep[e.g.][]{Pont2008, Pont2013, Sing2016}, photochemistry of long hydrocarbons has been proposed as a formation mechanism for these haze species, in analogy to the formation of hazes in the atmospheres of the Earth \citep{Jacobson2000} and the Solar System giant planets \citep[e.g.][]{Yung1984, Gladstone1996, Moses2000, Wong2003}. For this reason, photochemical models \citep{Liang2004, Zahnle2009_photo, Zahnle2016, Kawashima2018, Kawashima2019} and laboratory experiments \citep{He2018, Horst2018, Fleury2019, Fleury2020, Yu2021} have specifically targeted the formation and evolution of hydrocarbon hazes via photochemistry.

% what do we already know?
Previous 1D photochemical studies of irradiated exoplanets found that the upper layers ($p<10^{-3}$~bar) of hot Jupiters can be strongly affected by photochemistry, in particular through the photolysis (i.e.~dissociation after interaction with a photon) of molecules such as \chem{CH_4} and \chem{NH_3} \citep{Zahnle2009_photo, Moses2011, Venot2012, Rimmer2016, Drummond2016, Hobbs2019}. The resulting large amounts of H \citep{Liang2003} and other atomic components \citep{Moses2014} are then free to diffuse and recombine into different species, such as \chem{HCN} and \chem{C_2H_2}. Photochemical kinetics studies have demonstrated a greater effect of photochemistry -- and chemical disequilibrium in general -- in exoplanets with lower effective temperatures, both because cooler atmospheres have slower chemical timescales, as well as high abundances of photochemically active species \citep{Moses2014}.
The above photochemistry findings, established with hot Jupiter models, are likewise supported by hydrogen-rich sub-Neptune models \citep{Miller-RicciKempton2012, Moses2013, Hu2014, Miguel2014}. However, the relative importance of photochemistry varies with modelling choices such as the thermal profile \citep{Venot2014}, elemental composition \citep{Line2011, Moses2013}, and vertical mixing efficiency \citep{Agundez2014_GJ436b, Miguel2014}.
Additionally, photochemical networks including sulfur have been used to infer relatively high abundances of \chem{H_2S} and \chem{HS}, as well as abundance discrepancies of non-sulfurous species relative to pure C/N/O/H networks \citep{Zahnle2009_sulfur, Hobbs2021, Hu2021, Tsai2021}.
Other studies, still, have focused on the uppermost layers of the atmosphere, including ion photochemistry \citep{Yelle2004, GarciaMunoz2007, Koskinen2013, Lavvas2014, Rimmer2016, Helling2019_andRimmer, Barth2021, Locci2022}. Connecting the different physical regimes in the upper and lower atmospheres of exoplanets is still an active topic of research.

% complications and inadequacies of 1D models
Given the unique day-night dichotomy of tidally locked exoplanets, several two-dimensional \citep{Agundez2014, Venot2020_wasp43b, Moses2021, Baeyens2021} and three-dimensional \citep{Cooper2006, Mendonca2018_disequilibrium, Steinrueck2019, Drummond2018_metallicity, Drummond2018_HD209458b, Drummond2018_HD189733b, Drummond2020} models have been employed to investigate the spatial variations in the atmospheric chemistry of exoplanets. These studies have demonstrated the potential of efficient zonal \citep[and meridional,][]{Drummond2018_HD189733b, Drummond2020} advection to remove longitudinal chemical variations, primarily in cooler planets \citep{Baeyens2021}. The averaged chemical composition is then mostly similar to that of the day side. However, because of the computational cost of 3D chemistry-coupled GCMs, most studies have relied on parametrized chemistry rather than chemical kinetics, and as such have not included photochemistry.
The pseudo-2D studies that did include photochemical reactions \citep{Agundez2014, Venot2020_wasp43b, Moses2021} have found zonal variations down to $10^{-4}$~bar due to photodissociation, and a potentially large photochemical production of \chem{HCN}. Vigorous atmospheric mixing processes transport photodissociated day-side species to deeper layers, or to the night side of the planet, where they recombine. Cool planets with $\Teff=500-700$~K potentially experience the strongest impact of photochemistry \citep{Moses2021}.

The parameter space for photochemistry is nevertheless largely unexplored in a 2D context. \citet{Moses2021} have varied the effective temperature and metallicity of photochemical models, but they did not explicitly investigate the impact of the host star. The stellar spectrum, size, and distance, all influence the UV flux impinging on the planetary atmosphere and the strength of photochemical processes. Moreover, it remains unclear in which planets photochemistry could affect also the middle, photospheric region of the atmosphere ($\sim$1--100~mbar), where it could play a role in the formation of infrared transmission features.

% how we will solve these questions
In this paper, we aim to find out which planets experience large effects of photochemistry impacting their atmospheric composition, and which planets do not. To this end, we build upon the grid of pseudo-2D chemistry models developed in \cite{Baeyens2021}, by incorporating photochemical reactions, in addition to the previously included processes of vertical and horizontal mixing. We further synthesize transmission spectra for each model, estimating the extent to which photochemistry modifies the observations.

The outline of this paper is as follows. In Section~\ref{sec_phot_modelling}, the chemical kinetics code and the framework of the model grid are presented. Since we build upon the results of \cite{Baeyens2021}, we focus mostly on describing the implementation of photochemistry. The new photochemical models are presented in Section~\ref{sec_phot_results}, and discussed in a broader context in Section~\ref{sec_phot_discussion}. Specifically, we focus on whether photochemistry affects the deep atmosphere or the night side (Section~\ref{sec_phot_disc_deep}). We also discuss the impact of a hot thermosphere (Section~\ref{sec_thermosphere}), a comparison to past photochemical models (Section~\ref{sec_phot_disc_earlier}), and possible implications for high-resolution spectroscopy (Section~\ref{sec_phot_disc_highres}) and hazes (Section~\ref{sec_phot_disc_haze}). Finally, our conclusions are presented in Section~\ref{sec_phot_conclusion}.

\section{Modelling Tools}\label{sec_phot_modelling}

The chemical models are computed with the pseudo-2D chemical kinetics code developed by \citet{Agundez2014}, which includes disequilibrium chemistry via vertical mixing ($\kzz$), horizontal advection and photochemical reactions. %, i.e.~chemical reactions involving a photon.
We compute our grid of photochemical models in largely the same way as described in \cite{Baeyens2021}, namely by post-processing the outcomes of a 3D GCM grid with the chemical kinetics code. However, the inclusion of photochemistry requires some changes to the modelling setup, which are detailed below.

As in \cite{Baeyens2021}, we assume a solar elemental composition, which implies a C/O ratio of 0.55. The pseudo-2D code is initialized with a fully converged, one-dimensional photochemical kinetics model, corresponding to the substellar point. It is then integrated for a total of 50 rotations, so that a periodic steady-state can be reached.

\subsection{Upper Atmosphere Extension}\label{sec_extension}

Because photochemistry mainly occurs in the upper atmosphere, we have extended the pressure domain of the simulation from $10^{-4}$~bar \citep{Baeyens2021} to $10^{-8}$~bar. For some cold models, an upper boundary of $10^{-7}$~bar was chosen to avoid numerical issues. Although a $10^{-8}$--$10^{-7}$~bar upper boundary pressure is not unusual for chemical kinetics models \citep[e.g.][]{Agundez2014, Venot2020_wasp43b, Moses2021}, it is not a conservative choice, since it reaches well into the outer regions of the planet's atmosphere.\footnote{For comparison, in the Earth's atmosphere, a pressure of $10^{-8}$~bar is attained at altitudes between 100 and 150~km \citep{COESA1976_standard_atmosphere}. This is above the K\'arm\'an line of 100~km, the conventional boundary for outer space.} Accordingly, we have increased the number of vertical layers in the chemical model from 45 to up to 120.

The GCM models in our grid did not reach the low pressures up to $10^{-8}$~bar, so assumptions have been made to extrapolate the temperature- and $\kzz$-profiles, needed for the chemistry code, from $10^{-4}$~bar upward. Analytic calculations \citep{Guillot2010} and GCM simulations \citep[e.g.~Fig.~4,][]{Baeyens2021} alike suggest that the temperature becomes approximately isothermal at pressures below $10^{-2}$~bar. Therefore, we have taken the simplest approach: to extend the temperature profiles isothermally upward from their value at $10^{-4}$~bar.
It should nevertheless be noted that thermospheric models predict a strong temperature increase at pressures of $p < 10^{-6}$~bar \citep{Yelle2004, GarciaMunoz2007, Koskinen2013}, challenging the isothermal assumption made here. However, since we are mostly concerned with the impact of photochemistry on the atmospheric layers around 10--100~mbar that are probed with transmission spectroscopy, our isothermal extension should be sufficiently accurate. Still, the potential influence of thermospheric heating on our photochemical models is examined in Section~\ref{sec_thermosphere}.

Analogously, because we cannot derive the eddy diffusion coefficient at high altitudes from a GCM simulation, we simply adopt the value of $\kzz$ at $10^{-4}$~bar \citep[see Fig.~7,][]{Baeyens2021} for lower pressures as well. Although $\kzz$ generally increases with height \citep{Parmentier2013, Komacek2019, Baeyens2021}, it is expected to reach approximately constant values in the upper atmosphere \citep[][and references therein]{Steinrueck2021, Moses2021}. Given the uncertainties to this parameter, we opt again for a simple approach in all models. Finally, a molecular diffusive term is also included in the model, although it is not expected to play an important role at pressures greater than $10^{-8}$~bar \citep{Venot2012}.

\subsection{Photochemical Network}\label{sec_network}

Another difference between the models computed in this work and the models presented in \citet{Baeyens2021}, is the chemical network. Here, we exchanged the reduced chemical network of \citet{Venot2020_network} for the full network. Using the full network of \citet{Venot2020_network} is required when including photochemistry, because the reduced network does not include photolysis reactions, nor is it optimized to accurately represent the upper atmosphere when photochemically active species are included. As such, the network we use here is the full network of \citet{Venot2020_network}, which includes 108 chemical species consisting of C, N, O, H, with 1906 reactions (948 of which are reversible, and 10 irreversible). The network also contains 52 photolysis reactions.

\begin{figure}
  \centering
  \includegraphics[width=\columnwidth]{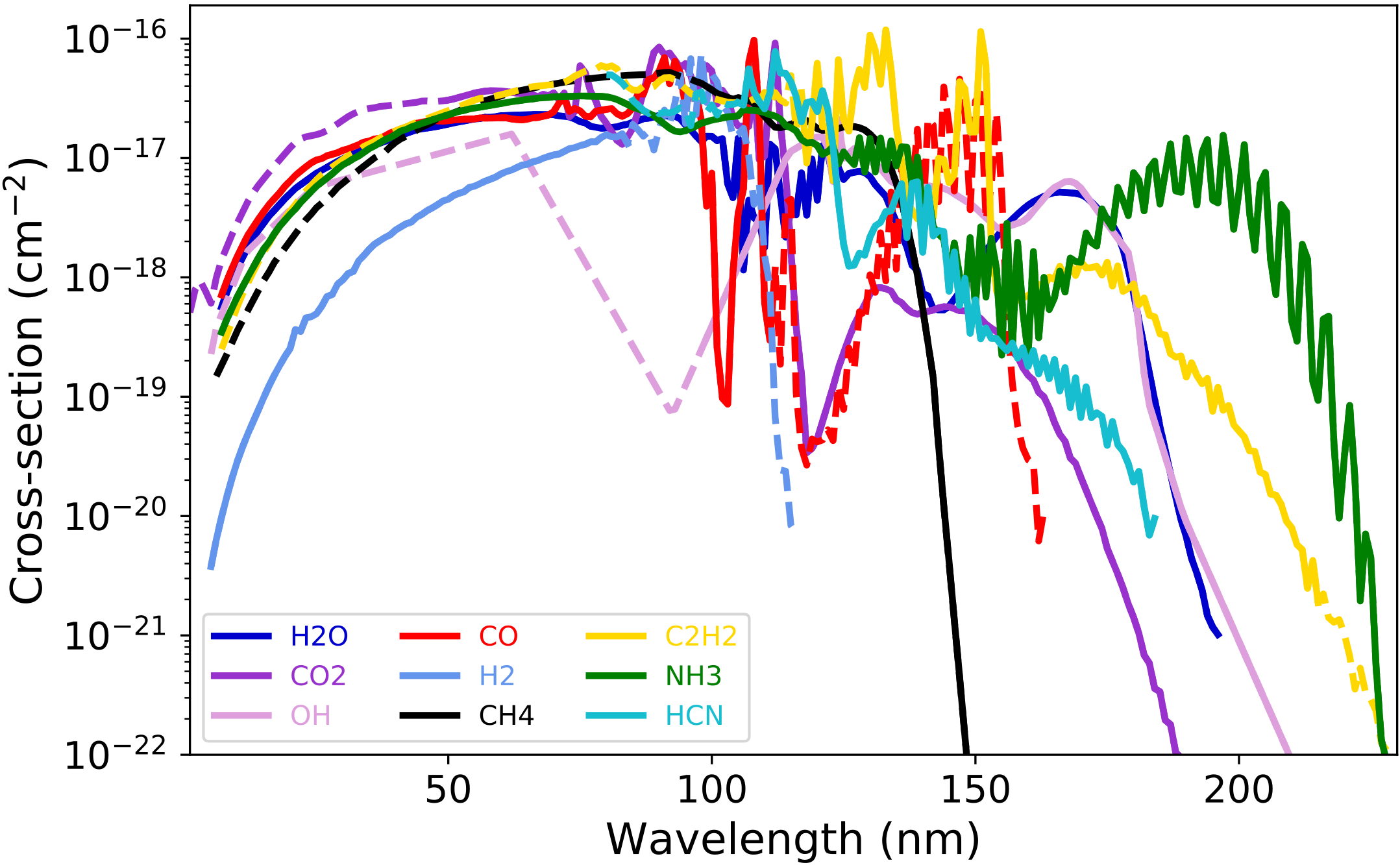}
  \caption[Photo-absorption and photodissociation cross-sections as a function of wavelength]{The photo-absorption (\emph{dashed lines}) and photodissociation (\emph{full lines}) cross-sections for a selection of abundant molecules in exoplanet atmospheres are highest in the UV wavelength range, and drop rapidly towards the visible wavelengths.}
  \label{fig_cross-sections}
\end{figure}

The photoabsorbing species are \chem{H_2O}, \chem{CO_2}, \chem{H_2CO}, \chem{OH}, \chem{OOH}, \chem{H_2O_2}, \chem{CO}, \chem{H_2}, \chem{HCO}, \chem{CH_3OH}, \chem{CH_3OOH}, \chem{CH_2CO}, \chem{CH_3CHO}, \chem{CH_3}, \chem{CH_4}, \chem{C_2H}, \chem{C_2H_2}, \chem{C_2H_3}, \chem{C_2H_4}, \chem{C_2H_6}, \chem{N_2}, \chem{NH_2}, \chem{NH_3}, \chem{N_2H_4}, \chem{HCN}, \chem{H_2CN}, \chem{C_2N_2}, \chem{NO}, \chem{NO_2}, \chem{NO_3}, \chem{HNO_2}, \chem{HNO_3}, \chem{N_2O}, \chem{N_2O_4}, and \chem{N_2O_3}. For each of these species, wavelength-dependent cross-sections and quantum yields are used to determine the probability of a molecule absorbing a photon, and the fraction of such absorption events that result in molecular dissociation. For many reactions, multiple dissociation pathways are possible. Almost all photoabsorption cross-sections reside in the (X)UV domain and fall off quickly for wavelengths larger than 200~nm (Fig.~\ref{fig_cross-sections}). Some species have cross-sections that reach into the blue visible light, and three of them (\chem{HCO}, \chem{NO_2}, and \chem{NO_3}) absorb in the full visible wavelength range. The photochemical data consist of mixed laboratory measurements and theoretical calculations, and are described in detail in \citet{Venot2012, Hebrard2013} and \citet{Dobrijevic2014}.
Although photochemical cross-sections are known to be temperature dependent \citep[e.g.][and references therein]{Venot2018}, the exact temperature-dependence is usually not well measured. Therefore, and since the effect of temperature-dependent cross-sections on transmission spectra is expected to be small \citep{Venot2018}, we do not take temperature-dependence of photochemical cross-sections into account here.

% \begin{table}
% 	\centering
% 	\caption{Overview of the species with photoabsorption and photodissociation reactions, with references.}
% 	\label{tab:photochem}
% 	\begin{tabular}{r r}
%     \hline
%     \hline
%     Species & Reference \\
%     \hline
%     \noalign{\smallskip} \\
%     \noalign{\smallskip}\hline
%     \end{tabular}
% \end{table}

\subsection{Stellar Spectra}\label{sec_stellar_spectra}

\begin{figure}
  \centering
  \includegraphics[width=\columnwidth]{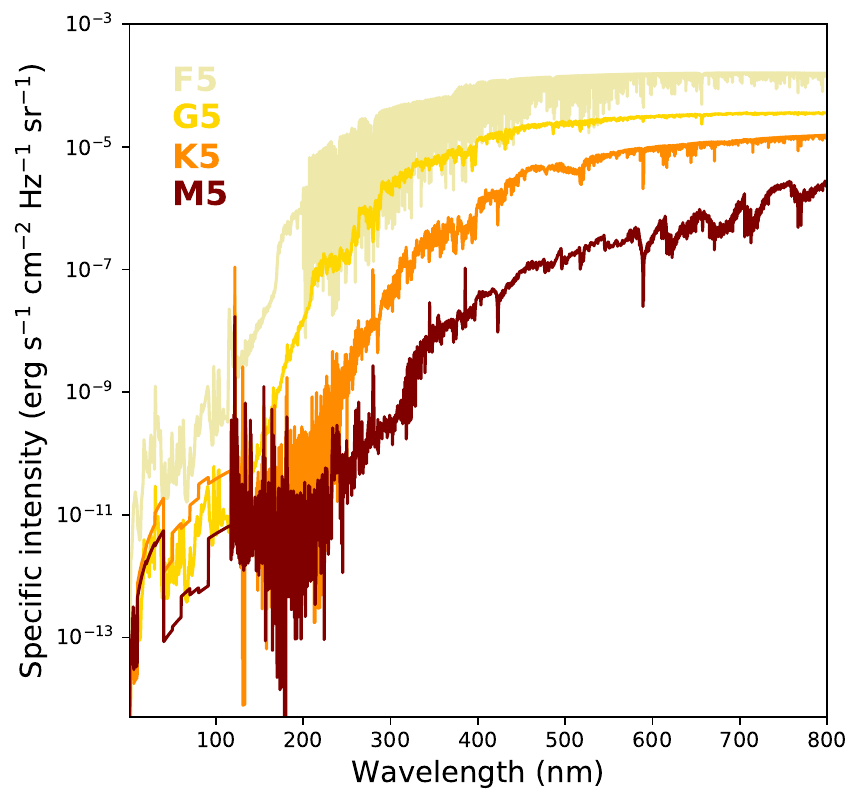}
  \caption[Spectral energy distributions used in the pseudo-2D chemistry code]{The spectral energy distributions (SEDs), used to compute the photodissociation rates, are shown for the different stellar types included in our grid.}
  \label{fig_stellar_spectra}
\end{figure}

To compute the photochemical dissociation rates, the incoming (UV) flux needs to be known. This quantity is provided via stellar spectral energy distributions, which are assumed to be constant in time. Given the framework of our grid, we adopt four spectral energy distributions ranging from 1~nm to 800~nm, which we apply for the M5, K5, G5, and F5 simulations (Fig.~\ref{fig_stellar_spectra}).

For the M- and K-star simulations, the spectral energy distributions are part of the \emph{MUSCLES}\footnote{Measurements of the Ultraviolet Spectral Characteristics of Low-mass Exoplanetary Systems} Treasury Survey \citep{France2016, Youngblood2016, Loyd2016}, a sample of panchromatic spectra for 11 low-mass exoplanet host stars that is made available via \emph{MAST}\footnote{Mikulski Archive for Space Telescopes, \url{https://archive.stsci.edu/prepds/muscles/}}. These spectra are composed of PHOENIX stellar atmosphere models \citep{Husser2013} in the visible wavelength range, \emph{Hubble Space Telescope} observations in the UV range (with spectral reconstructions of Lyman $\alpha$ and extreme UV \citep{Youngblood2016}), and \emph{Chandra} and \emph{XMM-Newton} observations with plasma emission models \citep{Smith2001} in the X-ray range. From this survey, we choose the star GJ~876 as a representative for the M5-star spectrum, and HD~85512 for the K5-spectrum (Fig.~\ref{fig_stellar_spectra}), since these two stars correspond well to the photospheric temperatures assumed in our grid \citep{France2016}. Flaring activity has been detected in the former star \citep{Walkowicz2008, Poppenhaeger2010, France2012}, but both spectral energy distributions have been constructed by making use of observations during low levels of activity \citep{Loyd2016}.

For the G5-star spectral energy distribution, we have adopted the present-day solar spectrum \citep{Claire2012}. The Sun provides a sufficiently close match to the 5650~K effective temperature of the paradigmatic G5-host star needed for our grid, with an excellent data quality (Fig.~\ref{fig_stellar_spectra}). However, it should be kept in mind that the present-day Sun is rather inactive compared to the stage when it was younger or to other G-type stars \citep[e.g.][]{Ribas2005}, so the adopted spectrum again represents a reasonable lower limit on the UV flux.

For the F-type star, we have composed an F5-spectrum from different sources, due to a lack of spectral energy distributions covering the wavelength range from the X-ray regime to visible light. First, for the optical part up to 200~nm, again a PHOENIX model atmosphere is employed, with $\Teff=6500$~K, log~$g=4.5$, and solar metallicity \citep{Husser2013}. In the UV range between 115 and 200~nm, we use the composite spectrum of the F-type star HD~128167, in units of flux, prepared by \citet{Segura2003} based on observations by the \emph{International Ultraviolet Explorer} satellite. Finally, in the shortest wavelength range, between 1 and 115~nm, we have scaled the solar spectrum in such a way that the integrated luminosity in this bandwidth attains $10^{30}$~erg/s. This is a realistic value for F-stars, although luminosities between 10$^{28}$ and $10^{31}$~erg/s seem possible \citep{Panzera1999, Wang2020_xray_luminosities, Johnstone2021}. For the luminosity scaling, a value of $1.35$ $R_\odot$ was used as F5-star radius \citep{Hubeny2014}. The resulting spectral energy distribution has the highest intensity of the four spectra used in this paper (Fig.~\ref{fig_stellar_spectra}).

\begin{figure}
  \centering
  \includegraphics[width=\columnwidth]{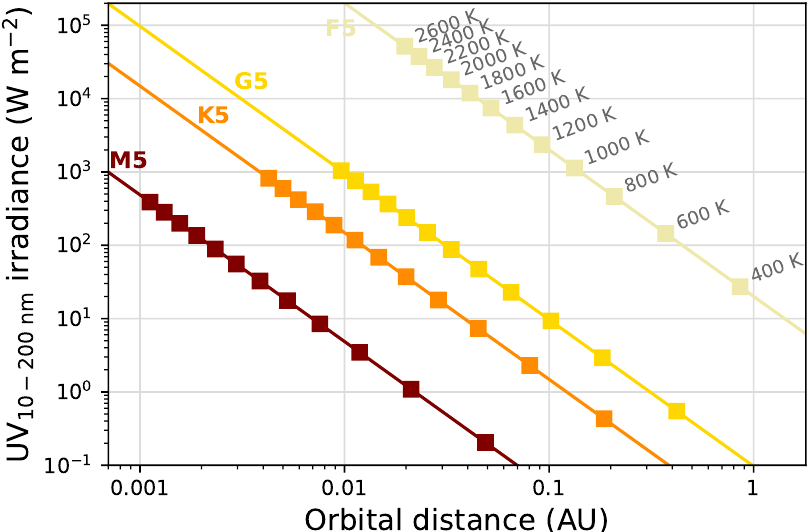}
  \caption[UV irradiance as a function of orbital distance]{The UV irradiance as computed for the wavelength range between 10~nm to 200~nm, based on the four different stellar spectra shown in Fig.~\ref{fig_stellar_spectra}. The UV irradiance is plotted as a function of the orbital distance. The models of our grid are situated on the figure with squares. For a given planetary temperature, the M-, K-, and G-type stars give rise to comparable UV irradiance; the difference being a factor of a few. Given our choice of spectra, the planets orbiting F-type stars can be considered relatively highly irradiated.}
  \label{fig_UV_irradiance}
\end{figure}

Because most parameters in our atmospheric model grid are self-consistently coupled under the assumption of synchronous rotation, each model is at a different orbital distance from its host star \citep[see Section~3 and Table~1 in][]{Baeyens2021}. The flux impinging on the top of the atmosphere depends on the stellar intensity, stellar radius and semi-major axis, and will thus also be different for each model. The stellar radii for each stellar type are adopted from \citet{Hubeny2014} (their table~2.4) and equal 0.27, 0.68, 0.91, and 1.35~$R_\odot$ for the M5-, K5-, G5-, and F5-stars respectively. In order to compare the expected rate of photochemical dissociations between models, we have computed the UV irradiance at the top of the atmosphere of each model (Fig.~\ref{fig_UV_irradiance}). It appears that for the M-, K-, and G-type hosts, the distance-dependence of the planetary effective temperature approximately compensates for changes in the stellar luminosity, so that planets with the same effective temperature experience comparable UV irradiance. The planets with an F-type host, on the other hand, see a consistently higher UV flux for the same temperature, in line with their comparatively high UV intensity (Fig.~\ref{fig_stellar_spectra}).

Finally, we note that the chromospheric emission, and equivalently the UV flux, of stars is correlated with stellar activity, as well as stellar parameters such as age and rotation \citep{Linsky2020}. Therefore, the scatter among stars is large, and it is not a trivial exercise to typify the X-ray and UV spectrum of a certain stellar type, nor is it evident to capture the diversity in the observed stellar population. Even within the small \emph{MUSCLES} sample, the estimated photodissociation rates can vary by an order of magnitude \citep{Loyd2016}. The scope of this work is to investigate the impact of photochemistry in a 2D-chemistry context for different planets. However, a future study investigating the chemical response to different parametrized stellar fluxes \citep[cfr.][]{Kawashima2019} would be beneficial.

\subsection{Model Runtimes}

We briefly discuss the computational cost of taking into account photochemical reactions in the chemical kinetics code. To quickly reach model convergence, we adopted the same approach as laid out in Appendix~A2 of \citet{Baeyens2021}, which consists of initializing the code with a converged 1D chemical kinetics model at the substellar point. Since the vertical advection timescales are slower than those associated with photochemical dissociations, we did not notice a difference in the simulation time needed for convergence when photochemical reactions are included. We note that our similar runtimes for the cases with and without photochemistry should not necessarily be viewed as a general result, since the model setup and exact implementation of photochemical reactions in the kinetics model dictate the computation time. 

Moreover, the necessary inclusion of a new chemical network with additional species and more reactions (Section~\ref{sec_network}) as well as an extension of the vertical grid (Section~\ref{sec_extension}) lead to considerably longer runtimes when photochemistry is included. Specifically, a simulation making use of the full chemical network was recorded to run up to 80 times slower than a simulation with the reduced network. Thus, future studies would do well to balance the need for accurate photochemistry with the computational constraints introduced by using the complete chemical network, especially regarding the implementation in 2D or 3D models \citep[e.g.][]{Drummond2020}.

\section{Simulation Results}\label{sec_phot_results}

\subsection{Chemical Composition}

\begin{figure*}
  \includegraphics[width=\textwidth]{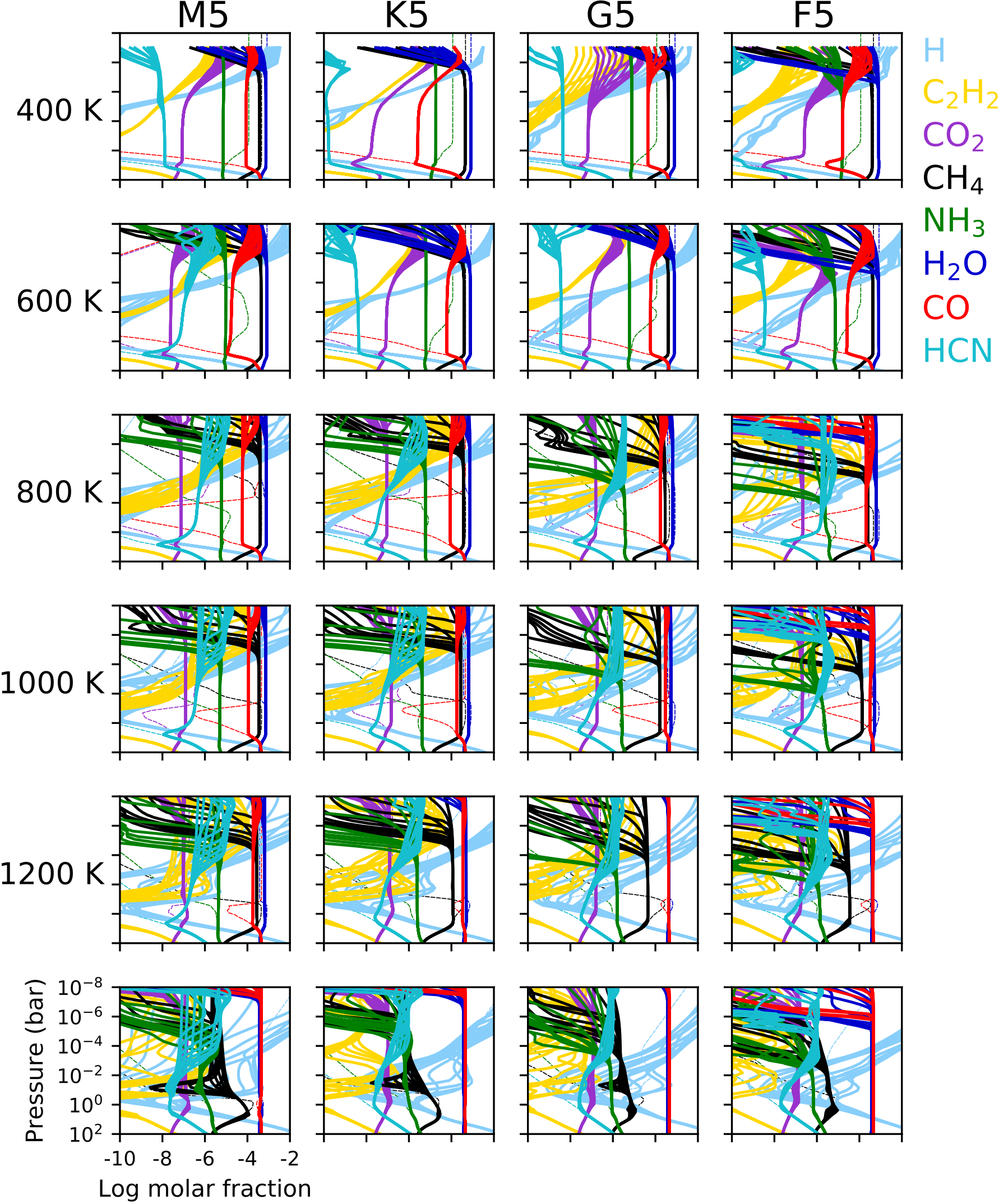}
  \caption[Grid of cool pseudo-2D photochemistry models]{The cool (400--1400~K) photochemical models in our grid show strong photolysis in the upper atmosphere, with substantial number fractions of acetylene (\chem{C_2H_2}). For each species (see colour legend), twelve profiles are plotted, corresponding to longitudes -180$^\circ$, -150$^\circ$, \ldots, 150$^\circ$. The dashed lines show the chemical equilibrium composition at the substellar point for comparison.}
  \label{fig_photochemistry_cool}
\end{figure*}

\begin{figure*}
  \includegraphics[width=\textwidth]{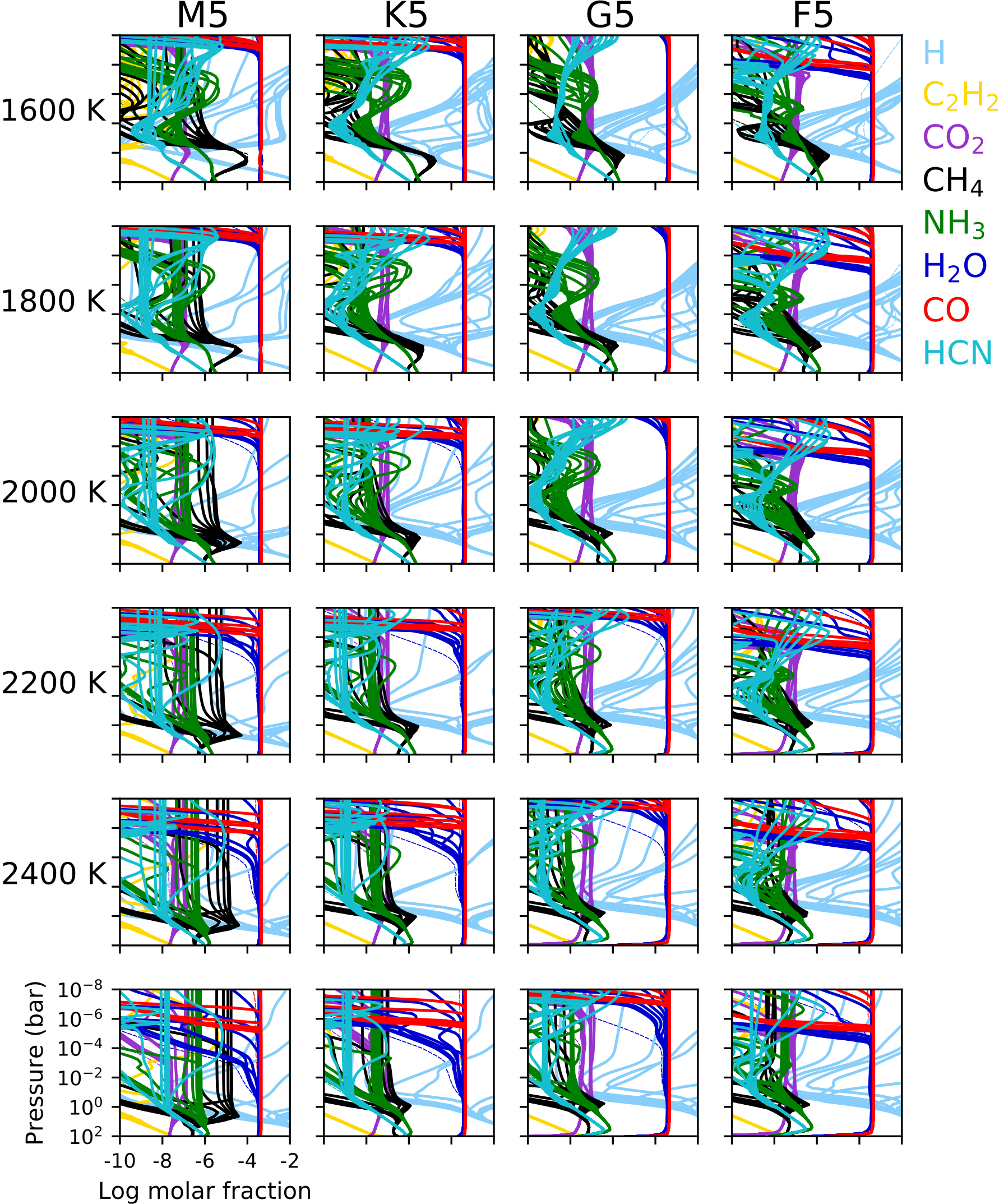}
  \caption[Grid of hot pseudo-2D photochemistry models]{The hot (1600--2600~K) photochemical models in our grid have molecular number fractions that vary strongly with pressure and longitude. For each species (see colour legend), twelve profiles are plotted, corresponding to longitudes -180$^\circ$, -150$^\circ$, \ldots, 150$^\circ$. The dashed lines show the chemical equilibrium composition at the substellar point for comparison.}
  \label{fig_photochemistry_hot}
\end{figure*}

Some of the most abundant molecular species, not counting \chem{H_2} and He, resulting from our pseudo-2D photochemical models with $g=10$~m/s$^2$, are shown as a function of pressure at different longitudes (Fig.~\ref{fig_photochemistry_cool} for $\Teff = 400$--1400~K; Fig.~\ref{fig_photochemistry_hot} for $\Teff = 1600$--2600~K). The influence of photochemistry can clearly be seen, especially in the upper atmosphere ($p< 10^{-4}$~bar), where the chemical composition for all temperatures is characterized by rapid drops in the abundances of some species, as well as large variations with longitude. Furthermore, some photochemically active species, such as \chem{HCN}, \chem{C_2H_2} or atomic hydrogen, display generally rising trends with altitude. Although the inclusion of photochemistry in our simulations modifies the chemical abundances strongly compared to a case with only mixing \citep[Fig.~8 in][]{Baeyens2021}, the general conclusions that have been established using models without photochemistry remain valid.


On comparing the outcomes of our models to the chemistry simulations that did not include photochemistry \citep[Fig.~8 in][]{Baeyens2021}, one can see that there is a good agreement in the abundances of species situated in the lower and middle atmospheres ($p > 1$~mbar). Chemical equilibrium and the quenching points are well reproduced, as we would indeed expect when switching between the reduced and full \citet{Venot2020_network}-networks. The dominant molecules in the atmosphere, not taking into account \chem{H_2} and \chem{He}, are still \chem{H_2O}, supplemented by \chem{CH_4} in the cold atmospheres ($\Teff \lesssim 1000$~K) and by \chem{CO} in the hot atmospheres ($\Teff \gtrsim 1000$~K). In the uppermost layers of the atmosphere ($p<10^{-6}$~bar), however, photolysis of \chem{CH_4} and \chem{H_2O}, can trigger \chem{CO} production, making it the most abundant carbon-species at high altitudes even in cold atmospheres (e.g.~at 600~K, Fig.~\ref{fig_photochemistry_cool}). Moreover, in the 400~K F-host model it can even be seen that photochemically produced CO and \chem{CO_2} are mixed downward in sufficient quantities such that the overall tropospheric abundance is raised by an order of magnitude above the quenched abundance. This is evidenced by abrupt variations where the vertical profiles of CO and \chem{CO_2} connect to their respective quench points near 10~bar.

The dichotomy identified by \cite{Baeyens2021}, in which cool atmospheres below 1400~K are zonally homogeneous, whereas warmer atmospheres are not, is still mostly valid, but only for pressures greater than 1~mbar. At lower pressures, photochemically driven day-night variations become noticeable, as the horizontal advection timescale is too slow to remove perturbations introduced by the increasing UV flux at the day side. Moreover, some photochemically produced species such as \chem{C_2H_2} or H vary zonally at even higher pressures.

An important consequence of the photodissociations of some species, is the increased amount of hydrogen cyanide (\chem{HCN}) and acetylene (\chem{C_2H_2}). Acetylene is more abundant in cool atmospheres, where it can reach a mixing ratio of $10^{-7}$ in the middle part ($\sim$10~mbar) and of $10^{-4}$ in the upper part (10$^{-4}$~bar) of the atmosphere. In simulations with $\Teff > 1400$~K, the acetylene abundance drops quickly, together with that of methane. \chem{HCN} reaches even higher abundances in the middle atmosphere, up to $10^{-6}$, as it is produced at altitudes above the quenching point from dissociated ammonia and methane, according to the net reaction \chem{NH_3} + \chem{CH_4} $\rightarrow$ \chem{HCN} + 3\chem{H_2} \citep{Moses2014}. At higher temperatures, the quenching points of \chem{HCN} and its parent species shift to lower pressures, resulting in a lower abundance in the mid-atmospheric region ($\sim$1--100~mbar). In the upper regions, even for hotter cases, appreciable day-side quantities of \chem{HCN} can potentially be retained (Fig.~\ref{fig_photochemistry_hot}).

A sweet spot for photochemistry, exemplified by high abundances of \chem{HCN} and \chem{C_2H_2} and photolysis of \chem{CH_4} and \chem{NH_3}, appears to exist between 800 and 1400~K. In these models, dynamical mixing brings the atmosphere out of equilibrium, resulting in relatively high abundances of parent species like ammonia and methane. At higher effective temperatures, a larger part of the atmosphere remains in chemical equilibrium, and parent species have lower concentrations, so photochemical products are less easily formed. At lower effective temperatures than $\sim800$~K, parent species are plentifully available, but the UV flux is lower, owing to the larger orbital separation, so photochemistry is less effective. The UV irradiance is not the only factor, however, because the F-host simulation of 400~K receives about the same UV irradiance as the G/K-host simulations of 1000~K (Fig.~\ref{fig_UV_irradiance}), without showing similarly high abundances of \chem{HCN} and \chem{C_2H_2}. Rather, in the 400~K F-host model, production of \chem{CO} and \chem{CO_2} is observed above the quench point, demonstrating that the effective temperature is still a major factor influencing the chemical composition \citep[see also][]{Moses2021}.

The host-star type's effect on the chemical composition was mainly attributed to rotation and heat redistribution in \cite{Baeyens2021}. Here, additionally the UV flux becomes an important parameter depending on the stellar host. In fact, a slight enhancement of photochemical processes is to be expected in our simulations as the stellar type changes from M to G, with a bigger jump for the F-types \citep[][and Fig.~\ref{fig_UV_irradiance}]{Miguel2014}. Such a trend seems to be present indeed, with the F-type models consistently showing signs of molecular dissociation at greater depths than the other simulations with equal effective temperature (Figs~\ref{fig_photochemistry_cool} and \ref{fig_photochemistry_hot}). However, due to the inherent scatter in stellar XUV fluxes (see also Section~\ref{sec_stellar_spectra}), this trend can be viewed as a demonstration of the possible diversity in, rather than a strictly monotonous relationship of stellar type with, irradiation-driven photochemistry.

% thermal dissociation vs photodissociation
Finally, in the hot atmosphere simulations, the strong UV flux causes large parts of the upper day-side atmosphere to be depleted in \chem{CO} and \chem{H_2O}. In our isothermal upper atmosphere, which may not be a good assumption for these ultra-hot models \citep[e.g.][]{Lothringer2018, Arcangeli2018, Parmentier2018} or in general (see Section~\ref{sec_thermosphere}), most water is destroyed at the day side at pressures lower than $10^{-4}$~bar, and \chem{CO} at pressures lower than $10^{-6}$~bar. This photolysis comes on top of the thermal dissociation that is expected in ultra-hot Jupiters \citep[e.g.][]{Parmentier2018}. At the night side, on the other hand, the atomic constituents recombine into \chem{CO} and \chem{H_2O}. As such, photolysis of \chem{H_2O} could exacerbate thermally driven limb asymmetries in ultra-hot exoplanets. Furthermore, molecular dissociation and recombination have been shown to affect the planetary heat redistribution \citep{Bell2018, Tan2019, Roth2021} and potentially bias retrievals \citep{Pluriel2020}. It could therefore be worthwhile to take into account both thermal and photochemical dissociations when studying ultra-hot Jupiters, as was already indicated by \citet{Shulyak2020}.

\begin{figure*}
    \centering
    \includegraphics[width=\textwidth]{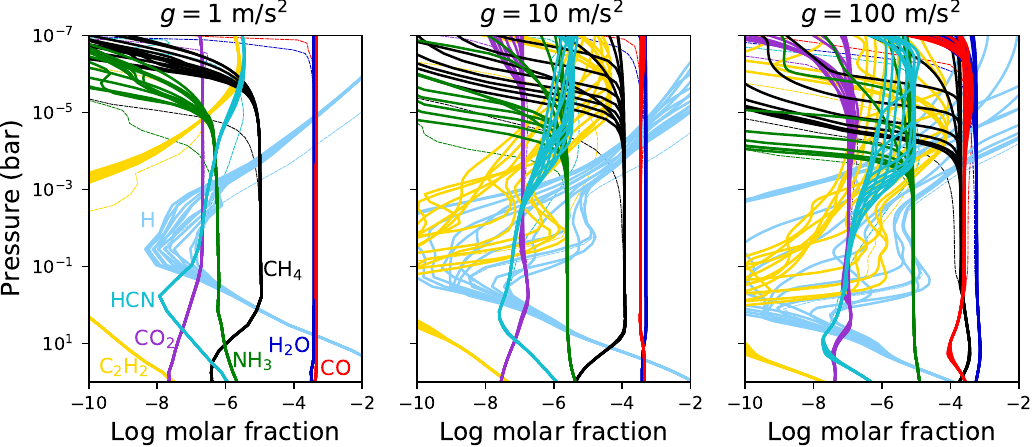}
    \caption{Upon comparing the chemical composition of a 1200~K-simulation with K5-host star and three different values for the gravity: 1~m/s$^2$ (\textit{left}), 10~m/s$^2$ (\textit{middle}), and 100~m/s$^2$ (\textit{right}), we find that a higher gravity results in a generally increased impact of photochemistry. This is exemplified by dissociation of \chem{CH_4} and \chem{NH_3} at higher pressures, and increased horizontal gradients of H and \chem{C_2H_2}. Chemical species are plotted at longitudes -180$^\circ$, -150$^\circ$, \ldots, 150$^\circ$, and colour coded the same way as Figs~\ref{fig_photochemistry_cool} and \ref{fig_photochemistry_hot}.}
    \label{fig_gravities}
\end{figure*}

\subsection{The Effect of Gravity}\label{sec_gravity}

The influence of gravity on two-dimensional photochemistry is investigated by running pseudo-2D chemical kinetics models corresponding to three different values of the planet's gravity, namely 1~m/s$^2$, 10~m/s$^2$, and 100~m/s$^2$. The other parameters of the simulation are an effective temperature of 1200~K and a K5-star host. The configuration with 10~m/s$^2$ is comparable to the hot Jupiter \hdeighteen{}. The necessary input data, such as temperature structures and vertical mixing efficiencies, are adopted from the GCMs of \citet{Baeyens2021}. As such, they are being self-consistently varied together with the gravity in order to obtain physically realistic simulations. In the models of \citet{Baeyens2021}, the gravity is changed by varying the mass while keeping the radius fixed. 

The three models with different gravity show clear changes in the pressure level that corresponds to the onset of photochemical dissociation (Fig.~\ref{fig_gravities}). In the low-gravity model, ammonia and methane get dissociated at pressures of 10$^{-5}$~bar or lower. In the high-gravity case, on the other hand, photodissociation already affects the chemical composition at pressures slightly lower than 1~mbar. These differences are partly due to the underlying discrepancies in thermal structure and mixing efficiency. Additionally, this phenomenon is explained when considering the optical depth as a function of pressure $\tau_\nu(p)$ in a hydrostatic atmosphere. This quantity is computed as 
\begin{equation}\label{tau}
    \tau_\nu(p) = \int^0_p{\frac{\kappa_\nu}{g} \textrm{d}p'},
\end{equation} where $\kappa_\nu$ is the opacity per unit mass. We note that, since the opacity depends on the chemical composition and thermodynamic state of the atmosphere, it also depends on the gravity itself. So equation \eqref{tau} is weakly non-linear in gravity. Nevertheless, for comparable atmospheric structures with different gravity, the factor $g$ in \eqref{tau} dominates \citep[e.g. see][]{Fortney2018}.

Thus, for a high-gravity planet, the optical depth becomes comparatively small and radiation will be able to penetrate to larger pressures. As such, high-gravity planets are more affected by photochemical processes. Given the relatively high gravity of \hdeighteen{} itself ($g=22$~m/s$^2$), it is conceivable that photochemical dissociation plays a role in the non-detection of methane in this planet via high-resolution spectroscopy \citep{Brogi2018}. However, a more tailored modelling effort of the high-resolution spectrum of \hdeighteen{} would be necessary to verify this hypothesis.

An additional result of the increased photochemistry at high pressures in the case of high gravity are the elevated abundances of acetylene and atomic hydrogen. These two species both attain molar fractions higher than 10$^{-6}$ in the 1 -- 100~mbar range, on the day-side of medium- and high-gravity planets (Fig.~\ref{fig_gravities}). In these cases, acetylene could be sufficiently abundant to enable its detection with the \textit{James Webb Space Telescope} \citep{Gasman2022}. The low-gravity model, however, only contains appreciable fractional abundances of \chem{C_2H_2} and H in the upper part of the atmosphere ($p < 1$~mbar). 

\subsection{Synthetic Spectrum}\label{sec_transmission_photochemistry}

To quantify the observational impact of photochemistry, we have computed synthetic transmission spectra for each model with the radiative transfer code \emph{petitRADTRANS} \citep{Molliere2019}. We have used the same approach as in \cite{Baeyens2021}, namely computing the transmission spectrum for the morning and evening limbs separately, and then averaging the squared results. The line opacities used in the spectral synthesis are \chem{H_2}, \chem{C_2H_2} \citep{Rothman2013_HITRAN}, O, \chem{O_2} \citep{Gordon2017_HITRAN}, \chem{C_2H_4} \citep{Mant2018}, \chem{H_2O}, \chem{CO}, \chem{CO_2}, \chem{OH} \citep{Rothman2010_HITEMP}, \chem{CH_4} \citep{Yurchenko2014_CH4}, \chem{NH_3} \citep{Yurchenko2011_NH3} and \chem{HCN} \citep{Harris2006_HCN, Barber2014_HCN}. For additional details, we refer to \citet{Baeyens2021} for the model setup, and \citet{Molliere2019} for the code. 

\begin{figure*}
  \centering
  \includegraphics[width=\textwidth]{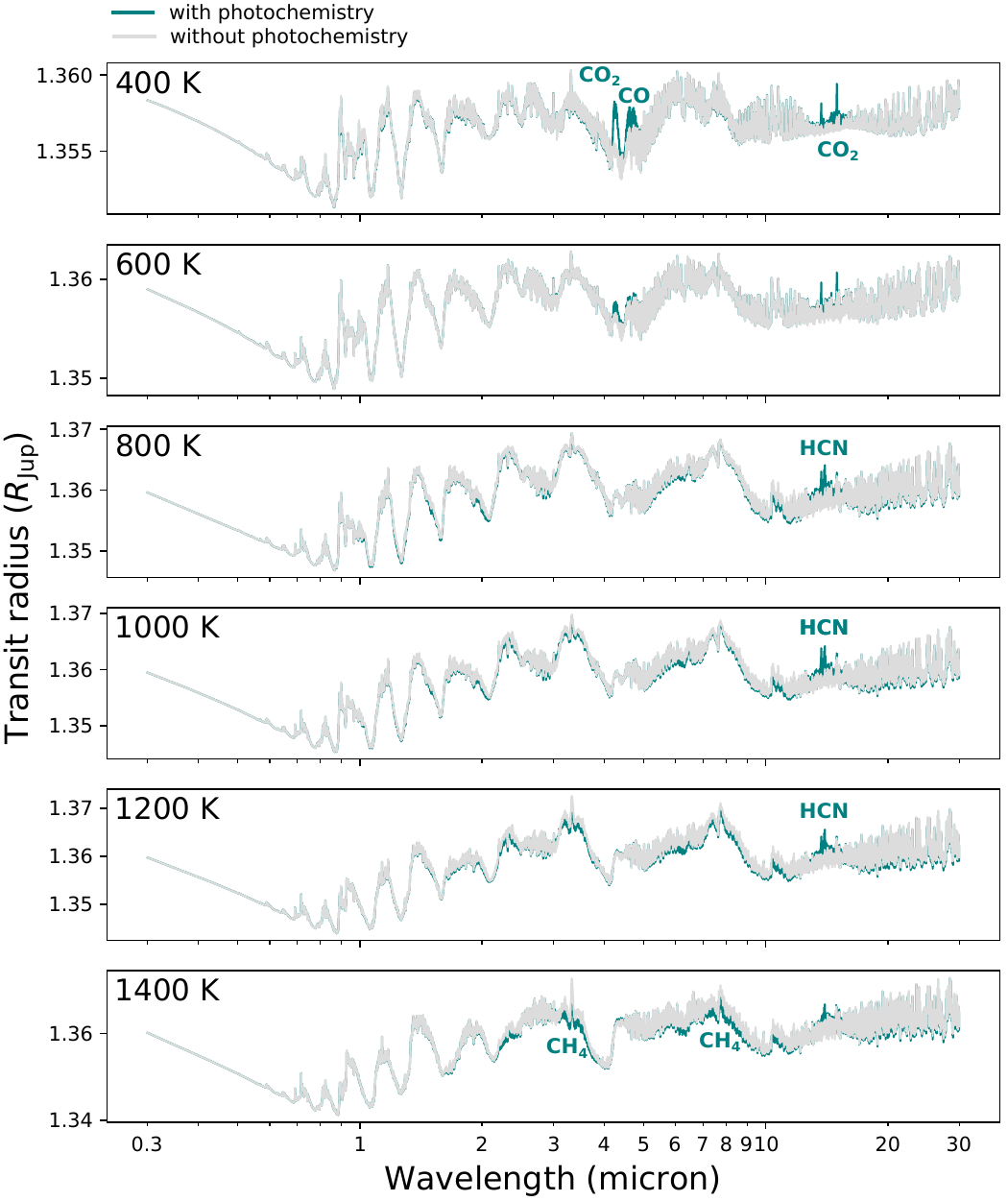}
  \caption[Synthetic transmission spectra with and without photochemistry]{The synthetic transmission spectra computed with \emph{petitRADTRANS}, corresponding to cool atmospheres with F-star hosts, show only small differences when photochemistry is included (\emph{teal}) or not (\emph{light grey}). Still, around 14~micron, increased \chem{HCN} absorption can be seen, as well as \chem{CO} and \chem{CO_2} absorption in the coldest models. In the warm ($\sim$1400~K) models, a reduced opacity caused by \chem{CH_4}-dissociation is noticeable as well. Note that the vertical axes are scaled differently.}
  \label{fig_transmission_photochemistry}
\end{figure*}

The effect of photochemistry on transmission spectra is relatively subtle for cool planets ($\Teff \leq 1400$~K, Fig.~\ref{fig_transmission_photochemistry}), and practically negligible for hot planets (not shown here), when low-to-medium resolution transmission spectroscopy from the \emph{Hubble} or \emph{James Webb Space Telescopes} is considered. Although we have shown that the impact of photochemistry on the chemical abundances is considerable, the transmission signature is formed near 10-100~mbar. These pressures are too high to experience the substantial photolysis of the upper atmosphere (Fig.~\ref{fig_photochemistry_cool}). The photochemical acetylene production at these pressures is probably also insufficient to be detectable in transmission.

\chem{HCN} thus appears to be the most promising observable signature of photochemistry. Especially in the effective temperature region of 800--1200~K, which we identified as a sweet spot for photochemistry earlier, \chem{HCN} shows a prominent absorption signature at 14~$\mu$m, in the same wavelength region where acetylene shows a strong opacity spike. The HCN spectral profile is also broader than the narrow \chem{C_2H_2} absorption. The signature of \chem{HCN}, potentially combined with acetylene, can thus result in transit depth differences of $\sim70$--100~ppm in our models with F- and G-star hosts (Fig.~\ref{fig_transmission_photochemistry}). The atmospheric chemistry for planets around different host stars is comparable (Figs~\ref{fig_photochemistry_cool} and \ref{fig_photochemistry_hot}), so a relatively small host star will be the most favourable to obtain a large transmission signature. Indeed, the difference in transit radius between the model spectra with and without photochemistry would result in a detectable transit depth difference of 1000~ppm for M-star hosts.

Another possible photochemistry signature arising from our synthetic transmission spectra, are the surprisingly large quantities of CO and \chem{CO_2} present in the 400~K models. This results in prominent opacity bumps for both species in the $4.3$--$4.6$~$\mu$m wavelength range, as well as around $15$~$\mu$m for \chem{CO_2}. In our models, photochemical enhancements of CO and \chem{CO_2} can give rise to a transit depth that is up to 100~ppm larger than when photochemistry is ignored. Photochemistry of CO and \chem{CO_2} has been studied before \citep{Line2011, Moses2014}, but not for effective temperatures as cold as 400~K and for strong UV irradiance.

Finally, we note that photochemistry can give rise to a decrease in \chem{CH_4} through photolysis in the 1200--1400~K range (Fig.~\ref{fig_transmission_photochemistry}), as well as modest middle-atmosphere enhancements of \chem{NH_3} in the 1600--1800~K range (Fig.~\ref{fig_photochemistry_hot}). Our synthetic observations suggest that such differences could give rise to 50--150~ppm changes in transmission spectroscopy. However, given the broad transmission signatures of methane and ammonia, we argue that such photochemical influences would be very degenerate with changes to the metallicity, C/O ratio, or $\kzz$, all of which can affect the abundances of methane and ammonia in this temperature regime \citep[see e.g.][]{Venot2014}. A higher C/O ratio in particular would result in more HCN and \chem{C_2H_2} production \citep{Venot2015_carbonrich}, similar to what is expected from photochemistry.  

We note that in this work we have focused on the analysis of exoplanet observations using transmission spectra rather than emission spectra or phase curves \citep{Moses2021}, because of two reasons. First, the climate models underlying our chemistry simulations have been performed using parametrized radiative forcing. Thus, the climate models lack a self-consistent energy budget, making this method not ideal for the quantitative synthesis of emission spectra \citep[see also the discussions on Newtonian cooling in][]{Showman2020, Baeyens2021, Schneider2022}. Second, we expect the effect of photochemistry to be more prominent in transmission spectra, because generally higher altitudes are probed (10 -- 100~mbar, compared to 10~mbar -- 10~bar for emission spectra). While it could be conceived that emission spectra at eclipse phase are especially sensitive to photochemistry, given their origin from the planetary day-side, we expect efficient horizontal mixing to skew the whole atmosphere toward the day-side composition anyway \citep[see][and Section~\ref{sec_phot_disc_deep}]{Agundez2014, Baeyens2021}.

\section{Discussion}\label{sec_phot_discussion}

\subsection{Photochemistry in the Deep Atmosphere and Night Side}\label{sec_phot_disc_deep}

That photochemistry affects the upper atmospheric day side, is evident. However, dynamical mixing processes can transport products of photolysis to the deep atmosphere, or to the night side, thereby affecting the chemical composition as a whole. By subtracting the abundances of simulations with and without photochemistry, but using an otherwise equivalent setup, the wider impact of day-side irradiance on the planetary atmosphere is gauged.

\begin{figure*}
  \centering
  \includegraphics[width=\textwidth]{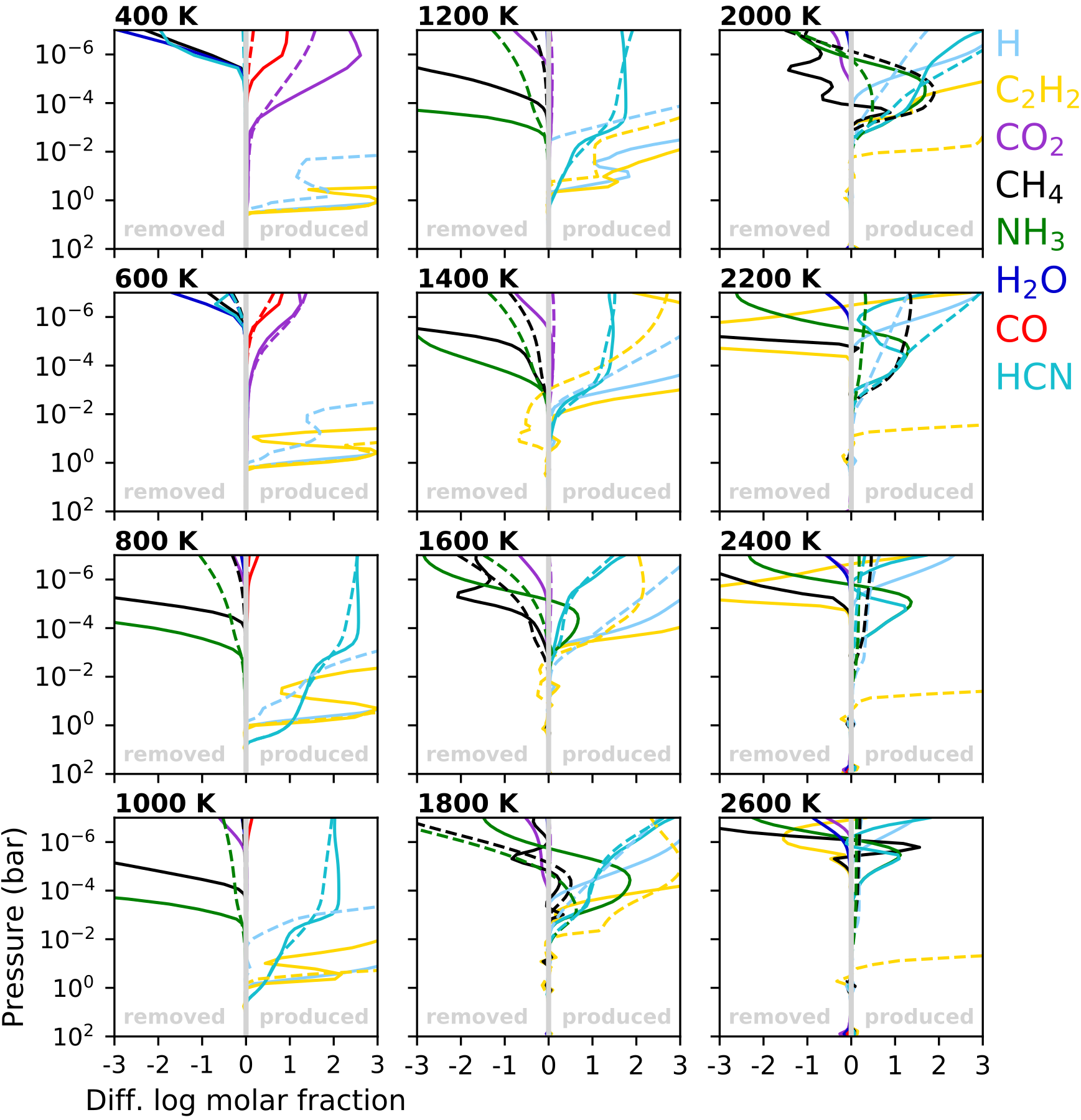}
  \caption[Differential abundances for models with and without photochemistry]{The difference between pseudo-2D photochemical models with photochemistry and without photochemistry, for the cases with $g=10$~m/s$^2$ and a G5-host, indicates the depth up to which photochemical processes change the atmospheric composition. For each species, the differential abundance at substellar (\emph{full}) and antistellar point (\emph{dashed}) are shown. A positive difference indicates a higher abundance of a species in the photochemical kinetics model.}
  \label{fig_diff_photo_onlymixing}
\end{figure*}

The depth at which photochemistry starts to have a strong impact on the chemical composition, appears to decrease with effective temperature (Fig.~\ref{fig_diff_photo_onlymixing}). In cold models, atmospheric layers at 4~bar already experience strongly increased acetylene and atomic hydrogen concentrations. This maximal pressure of influence for photochemistry is reduced to $10^{-2}$~bar for intermediate temperatures ($\Teff=1600$~K), and even $10^{-4}$~bar
for the hottest case, ignoring the consistent night-side enhancement of acetylene here\footnote{Note that, although the night side of hot photochemical simulations appears strongly enhanced in \chem{C_2H_2}, its concentration is still very low in absolute terms. In the 2600~K model, the molar fraction of \chem{C_2H_2} increases from 10$^{-22}$ to 10$^{-14}$ when photochemistry is included.}. Although the UV irradiance increases strongly with effective temperature (Fig.~\ref{fig_UV_irradiance}), it appears that the high day-side temperature of hot atmospheres gives rise to fast chemical reactions and efficient recombination of photolysed species. Hence, the atmosphere does not go out of chemical equilibrium.

Indirectly, photochemistry also affects the night-side composition, as the species that have been formed through photochemical processes on the day side of the planet are advected via the horizontal jet stream. Especially cool planets exhibit such efficient horizontal advection, as evidenced by the roughly coinciding abundances of photochemically produced species at the day side and night side of the planet (Fig.~\ref{fig_diff_photo_onlymixing}, H, \chem{C_2H_2}, CO, and \chem{CO_2} for $\Teff=400$--600~K; \chem{HCN} and \chem{C_2H_2} for $\Teff=800$--1600~K). The efficient day-night transport of photochemically produced species is in agreement with our finding that cool planets tend to be horizontally homogeneous, and globally evolve towards the day-side composition \citep{Baeyens2021}.

Species that are destroyed through photolysis on the day side also get advected towards the night side of the planet, resulting in moderate night-side depletions of ammonia and methane for effective temperatures between 800 and 1800~K (Fig.~\ref{fig_diff_photo_onlymixing}). However, in these cases, day-side depletion is much more pronounced, which indicates that, by the time methane- and ammonia-deficient air is advected, some photochemistry products have already been recycled into \chem{NH_3} and \chem{CH_4}. The partial replenishment of \chem{NH_3} and \chem{CH_4} after it has been destroyed through photolysis at the planetary day side can be attributed to vertical mixing. Indeed, an additional simulation without vertical mixing demonstrates that the night-side abundances of ammonia and methane remain low, despite them being thermochemically favoured here.
The gradual replenishment of photodissociated species, as gas is advected across the night side of the planet, further implies that the morning and evening terminator have different compositions, with the latter being more depleted in ammonia and methane (see also Fig.~\ref{fig_photochemistry_cool}).

Regarding exoplanet atmospheres with effective temperatures above 1800~K, our models suggest that photochemically enhanced or depleted species display discrepant concentrations on the day side and night side (Fig.~\ref{fig_diff_photo_onlymixing}), indicating that thermal effects dominate over horizontal advection. This is likely exacerbated when a temperature inversion or a hot thermosphere (Section~\ref{sec_thermosphere}) is taken into account.

We note that our model only takes into account horizontal advection in the eastward direction along the equator. Previous climate simulations \citep[e.g.][]{Rauscher2014_andKempton, Carone2020, Steinrueck2021, Baeyens2021} have demonstrated that the predominant wind direction at high altitudes is a direct day-to-night flow. Thus, it would be interesting to investigate whether such wind advection would increase the transport of photochemically produced species towards the night side. This would require a 3D chemistry-coupled GCM with photochemistry, which as of yet has not been developed for hot Jupiters.

\subsection{Upper Atmosphere Temperature}\label{sec_thermosphere}
% Does the upper atmosphere structure matter --> some experiments

Theoretical models for the upper atmospheres ($p < 10^{-6}$~bar) of hot Jupiters have suggested that strong irradiation and ionization cause the temperature to rise again, reaching values of up to 10000~K in the upper thermosphere \citep{Yelle2004, GarciaMunoz2007, Koskinen2013}. Similar trends have been confirmed by detections of the atomic line profiles of sodium, formed high in the atmospheres of hot Jupiters \citep{Vidal-Madjar2011, Huitson2012, Pino2018}. Therefore, our simple assumption of an isothermal extension of the pressure-temperature profiles is not expected to hold up to the upper boundary of $p = 10^{-8}$~bar. In order to gauge the sensitivity of our results to the upper atmospheric thermal structure, we investigate the chemical composition for a higher background temperature in the upper atmosphere.

We run our photochemical model again for typical hot Jupiter values ($\Teff=1400$~K, $g=10$~m/s$^2$, G5-star host), but with a hotter upper atmosphere. Instead of an isothermal extension for layers above $10^{-4}$~bar, we employ a parametrized thermosphere, constructed as follows. The pressure-temperature profiles are still derived from the GCM models in \citet{Baeyens2021}, and isothermally extended for pressures between $10^{-4}$--$10^{-6}$~bar, but for higher pressures we perform a logarithmic interpolation to an upper boundary temperature of 5000~K at $10^{-8}$~bar. This upper boundary temperature is further weighted with $\cos{\lambda}$, where $\lambda$ is the longitude with substellar point at $0\degrees$, so that only the day side is affected, and there is a smoother transition to the isothermal night side.
While this is a crude parametrization of thermospheric heating, it should be sufficient to investigate the impact of the upper atmosphere temperature structure on the chemical composition in the layers below.

\begin{figure*}
  \centering
  \includegraphics[width=\textwidth]{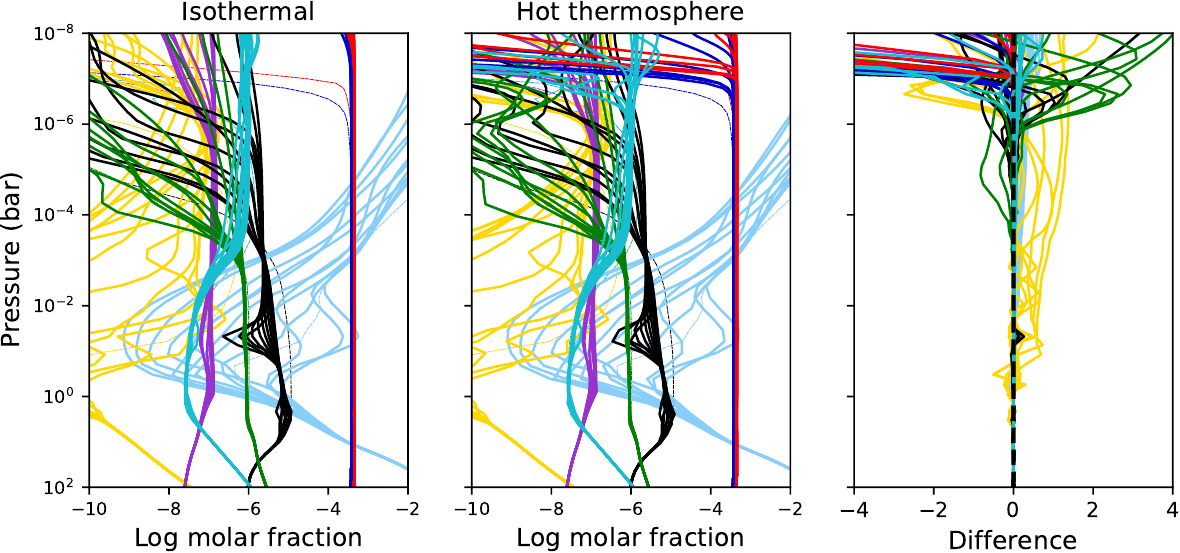}
  \caption[Chemical composition in the case of a hot thermosphere]{Our photochemical model ($\Teff = 1400$~K, $g=10$~m/s$^2$, and G5-host) that has an isothermally extended upper atmosphere (\emph{left}), and an equivalent photochemical model that has a hot thermosphere (\emph{middle}), are subtracted to illustrate the abundance differences (\emph{right}). The major differences are restricted to the upper atmosphere at pressures lower than 10$^{-6}$~bar. Chemical species are plotted at longitudes -180$^\circ$, -150$^\circ$, \ldots, 150$^\circ$, and colour coded the same way as Figs~\ref{fig_photochemistry_cool} and \ref{fig_photochemistry_hot}.}
  \label{fig_thermosphere}
\end{figure*}

From a comparison between a 1400~K-model with an isothermal photosphere, and one with parametrized thermospheric heating, we find that differences are generally restricted to pressures lower than 10$^{-6}$~bar, i.e.~the region where the heating is implemented (Fig.~\ref{fig_thermosphere}). Most notably different, the model with a hot thermosphere displays strong thermal dissociation of CO, \chem{HCN}, and water in the uppermost layers at the day side of the atmosphere. However, at the night side, where no thermospheric heating is assumed, the chemical composition is unaffected. We notice a small effect of a hot thermosphere on lower lying layers, namely a slightly higher acetylene abundance, mostly on the night side of the planet (Fig.~\ref{fig_thermosphere}). In absolute values, however, the acetylene concentration is not strongly affected.

Thus, we conclude that thermospheric heating only has a minor influence on composition at intermediate or high atmospheric pressures ($p < 10^{-4}$~bar). The regions where the heating takes place ($p<10^{-6}$~bar), however, are markedly changed. Hence, our simple assumption of an isothermal extension of the pressure-temperature profiles into the upper atmosphere is valid for the purposes of this study. A similar conclusion was reached by \citet{Moses2011}, and we confirm it here in a pseudo-2D framework.

\subsection{Comparison to Earlier Works}\label{sec_phot_disc_earlier}

In their one-dimensional chemical kinetics study, \citet{Moses2011} raised the question whether horizontal quenching could lead to the reconversion of photochemically produced species (\chem{HCN}, \chem{C_2H_2}, and \chem{H}) back into their parent species (\chem{NH_3}, \chem{CH_4}, and \chem{H_2} respectively) on the terminators or night side of irradiated planets.
Using pseudo-2D models including photochemistry, it was found that \chem{HCN} remains stable in high quantities after advection to the night side in hot Jupiter atmospheres of 1200--1400~K \citep{Agundez2014, Venot2020_wasp43b}. In these models, HCN shows little to no horizontal variations, except at high altitudes ($p< 10^{-4}$~bar). Furthermore, a qualitatively similar conclusion was reached for a suite of Neptune-sized exoplanet models \citep{Moses2021}. These authors also examined different effective temperatures, and found that acetylene and \chem{HCN} remain horizontally quenched in cool atmospheres, up until $\sim$1100~K and $\sim$1500~K respectively. This is in good qualitative agreement with our findings of strong photochemical contamination of the night side at low effective temperatures, although we obtain more zonal variation in the abundance of acetylene. Additionally, our models demonstrate a higher zonal variation for planets around lower-mass stars (e.g.~compare \chem{HCN} in the 1400~K, M5-, K5-, G5-models, Fig.~\ref{fig_photochemistry_cool}), because planets around cooler stars rotate faster and experience a higher day-night temperature contrast, with slower and narrower horizontal wind jets \citep{Baeyens2021}. A comparatively high XUV irradiance, however, can counteract this effect (F5-models, Fig.~\ref{fig_photochemistry_cool}).

\citet{Shulyak2020} have examined the impact of stellar irradiation on the exoplanet atmospheric chemistry in a 1D context, and have found a strong impact of photochemistry in hot stars (A0, F0) relative to cool stars (G2, K0). The authors suggest that disequilibrium chemistry can be detected in transmission spectra of 1000~K planets, through enhanced CO, \chem{CO_2}, % \chem{NH_3},
and HCN concentrations, but becomes undetectable in models of 2000~K. In this work, with a finer effective temperature grid and pseudo-2D chemistry, we find that HCN is best observed in the 800--1200~K temperature range, in agreement with \citet{Shulyak2020}, although our HCN mixing ratio and transmission signal are somewhat lower. This can be attributed to a different thermal structure, $\kzz$-profile, or (for altitudes above $p=10^{-2}$~bar) reduced HCN concentrations at the terminators in the pseudo-2D configuration. The enhanced CO and \chem{CO_2} abundances found by \citet{Shulyak2020} have also been addressed by \citet{Blumenthal2018} as a good indicator of disequilibrium chemistry around 4.5~$\mu$m in the 700--800~K effective temperature range. The additional absorption appears to stem mostly from vertical mixing, as highlighted in \citet{Baeyens2021} (Section~5.4) and in atmospheric retrievals \citep{Kawashima2021}. Here, we additionally find a photochemical enhancement below effective temperatures of 600~K (Fig.~\ref{fig_transmission_photochemistry}), reconfirming the wavelength region around 4.5~$\mu$m as a spectral region of interest in future transmission spectroscopy studies, such as those with the \emph{James Webb Space Telescope} (\emph{JWST}).

Finally, \citet{Hu2021} has investigated in detail the chemistry occurring in cold to temperate ($\Teff < 300$~K) gas planets. 
Here, an effect of the stellar type was identified through a low ammonia mixing ratio in the middle atmosphere (0.1--10~mbar) of planets orbiting K- and G-type stars, whereas M-stars allow for a higher ammonia content. A possible explanation is the reduced UV radiation of M-stars compared to their hotter counterparts \citep{Hu2021}. We do not find such effect in our coldest, 400~K-models, which all yield similar \chem{NH_3} abundances in the middle atmosphere. An additional outcome of the models by \citet{Hu2021} are CO and \chem{CO_2} as the most important photochemical products in cold environments where photolysis of ammonia is not taking place. This is likewise seen in our 400~K- and 600~K-models (Fig.~\ref{fig_photochemistry_cool}), where photolysis of \chem{CH_4} and \chem{H_2O} results in increased CO abundances. If \chem{NH_3} gets dissociated, an increased HCN abundance can be found as well. Additionally, we have highlighted the potential impact of CO and \chem{CO_2} on transmission spectra of cold planets (Fig.~\ref{fig_transmission_photochemistry}).

We note that for most of the works discussed here direct quantitative comparisons with our models are not straightforward due to different input physics. Potential sources for discrepancies include the temperature structure, vertical mixing, condensation, horizontal advection, and irradiation power. Thus, benchmarking efforts between different chemical models would be useful to reveal quantitative disagreements between different modelling tools.

\subsection{Implications for High-Resolution Spectroscopy}\label{sec_phot_disc_highres}

In Section~\ref{sec_transmission_photochemistry}, we have demonstrated the effect of photochemistry on the transmission spectra of irradiated exoplanets, which turns out to be limited. The reason is that photochemistry mostly affects the upper atmosphere, whereas the transmission spectrum is formed in the middle atmosphere at about 10--100~mbar. The \chem{HCN}/\chem{C_2H_2} absorption peak around 14~$\mu$m is therefore one of the only signatures of photochemistry in transit spectroscopy. Unfortunately, this signal will be difficult to detect as well, as it falls outside of the wavelength range of the \emph{JWST}/MIRI low-resolution mode\footnote{Although the \emph{JWST}/MIRI medium-resolution observing mode (MRS) will have a wavelength range of up to 28~$\mu$m, this mode will not be suitable for transmission spectroscopy due to read-out constraints.}.

Potentially, high-resolution spectroscopy of exoplanet atmospheres \citep[e.g.][]{Snellen2010, Birkby2018} could be very valuable for providing constraints on photochemical processes, and vice-versa. The technique is namely sensitive to the mid- to upper atmosphere, depending on the strength of the spectral lines measured. Indeed, detections of \chem{HCN} indicate its presence in the upper atmosphere of the hot Jupiters \hdtwenty{} ($\Teff=1400$~K, G-type host) and \hdeighteen{} ($\Teff=1200$~K, K-type host) in relatively high amounts \citep{Hawker2018, Cabot2019}. These measurements suggest a lower limit of 10$^{-6.5}$ for the concentration of HCN. Such a limit is compatible with our pseudo-2D models at around 1~mbar or lower pressures (Fig.~\ref{fig_photochemistry_cool}), assuming a solar metallicity and C/O ratio. However, additional detected molecules in the atmosphere of \hdtwenty{} have provided evidence of a high C/O ratio \citep{Giacobbe2021}. Likewise, a depletion of water with respect to CO in another hot Jupiter, has led to the same conclusion of a high C/O ratio \citep{Pelletier2021}. Both of these studies rely on the assumption of chemical equilibrium to infer high C/O ratios. However, we have shown here and in \citet{Baeyens2021} how disequilibrium chemistry can affect the atmospheric composition. In particular, photolysis of water at high altitudes could potentially play a role in explaining the water depletion. This motivates further investigation expanding the parameter space of chemical kinetics models to different metallicities and C/O ratios \citep[see also][]{Moses2013_COratio, Moses2021}.

Recently, it has also become possible for high-resolution spectroscopy to probe thermal and chemical gradients in exoplanet atmospheres, such as day-night or limb asymmetries \citep{Ehrenreich2020, Kesseli2021, Wardenier2021}. Such effects cannot be adequately captured via one-dimensional photochemical kinetics. Thus, 2D or 3D models that include photochemistry will remain uniquely useful tools in the interpretation and prediction of the spatial distribution of molecular species at high altitudes.

\subsection{Acetylene and Haze Formation}\label{sec_phot_disc_haze}

Acetylene (\chem{C_2H_2}) is a commonly detected molecule in the Solar System gas giants, with an abundance following the solar insolation \citep[e.g.][]{Guerlet2009, Giles2021}. Recent development of a high-temperature line list \citep{Chubb2020} has enabled its detection in a hot Jupiter atmosphere as well \citep{Giacobbe2021}. We identified a sweet spot of acetylene formation in exoplanet atmospheres between 800 and 1400~K (Section~\ref{sec_phot_results}). In this temperature range, acetylene can attain molar fractions higher than $10^{-7}$ in the mid- to upper atmosphere ($p < 10$~mbar). At these molar fractions, it could be detected with retrieval modelling of \textit{JWST} data \citep{Gasman2022}. As mentioned in Section~\ref{sec_transmission_photochemistry}, efforts of detecting acetylene with transmission spectroscopy will likely be hampered, however, due to the decreasing abundance of acetylene at higher pressures. 

With acetylene being a photochemical haze precursor molecule \citep{Zahnle2009_photo}, we hypothesize that the 800--1400~K temperature region could also be a sweet spot for the efficient formation of photochemical hazes in exoplanet atmospheres. \citet{Gao2020} have predicted that hydrocarbon hazes could be the dominant aerosol opacity source in the atmospheres of planets with equilibrium temperatures below 950~K (down to 700~K, the lower limit in their study). They find that haze formation becomes efficient when the atmosphere transitions from CO-dominated to \chem{CH_4}-dominated, a requirement that has been formulated before by \citet{Liang2004}. The transition temperature of 950~K is based on thermochemical equilibrium models. However, we have shown that appreciable methane concentrations can persist up until 1400~K with disequilibrium chemistry (Fig.~\ref{fig_photochemistry_cool}), so potentially, photochemical hazes follow this trend. Indeed, photochemical microphysics models \citep{Lavvas2017, Helling2020} have demonstrated the potential for haze formation in \hdeighteen{} and WASP-43~b, which have equilibrium temperatures of 1200~K and 1400~K respectively. Thus, even though haze production is expected to slow down at higher temperatures \citep{Kawashima2019}, it appears that the relatively efficient acetylene production in our photochemical models corresponds to the temperature range in which haze formation can be expected.

Finally, we note that laboratory experiments at 800~K have demonstrated that photochemical haze formation does not necessitate methane photolysis \citep{He2020}, and depends strongly on the initial mixture and atmospheric influences. Additional physics, such as ionizing radiation of stellar or cosmic origin, may also enhance the hydrocarbon concentration and facilitate haze production \citep{Rimmer2014, Barth2021}. It is clear that more sophisticated models, coupling disequilibrium chemistry, microphysics and radiative feedback, are necessary to pin down the exact dependencies of photochemical haze formation in exoplanet atmospheres.

\section{Conclusions}\label{sec_phot_conclusion}

In this study, we have investigated photochemistry in a suite of irradiated exoplanet atmospheres in order to gauge its impact on the atmospheric composition and on transmission spectra. To achieve this, we have used the pseudo-2D chemical kinetics code of \citet{Agundez2014} with the full chemical network of \citet{Venot2020_network}, and updated our grid of models for irradiated exoplanet atmospheres by extending them to higher altitudes and including photochemistry. 

In our photochemical models, we find that the upper atmospheres are depleted in \chem{CH_4} and \chem{NH_3}, and for high UV irradiance also \chem{H_2O} and \chem{CO}. On the other hand, in the middle atmosphere, \chem{HCN} and \chem{C_2H_2} are being produced from photolysed ammonia and methane in cool planetary atmospheres. The maximal pressure at which photochemical processes affect the chemical composition, decreases with effective temperature, and is about 4~bar in the coolest planet. Furthermore, we identify a sweet spot for the photochemical production of \chem{C_2H_2} and HCN between 800~K and 1400~K, and discuss the potential for photochemical haze formation in this temperature region.

Our models suggest an efficient longitudinal transport for effective temperatures below 1800~K, with a strong contamination of the night side by photochemically produced species from the day side, in agreement with other pseudo-2D models \citep{Agundez2014, Venot2020_wasp43b, Moses2021}. Photolysed ammonia and methane, however, are partially replenished on their way to the night side, suggesting a zonal chemical gradient with limb asymmetries. 

The impact of the host star on the chemical composition asserts itself mostly through the changing UV irradiance. In our models, photochemical processes are enhanced when the host star temperature increases, in agreement with \citet{Shulyak2020}. However, we note that this trend not necessarily applies universally, due to the large intrinsic scatter in stellar X-ray and UV luminosities \citep{Linsky2020}. Hence, future work continuing to quantify the spread of XUV luminosities of exoplanet host stars, as well as the corresponding variations that can be expected for planetary atmospheric photochemistry, would be very beneficial.

Despite the considerable impact of photochemistry on the atmospheric composition, the impact on transmission spectra was found to be feeble for planets with effective temperatures below 1400~K, and nearly negligible for planets with higher effective temperatures. The most prominent signature of photochemistry in transmission, is the enhanced absorption at 14~$\mu$m, but it will be difficult to detect, as it lies outside the wavelength range of the \emph{JWST}/MIRI stable, spectrophotometric observing mode. Potentially, a 50 to 150~ppm variation because of photochemically reduced methane or enhanced ammonia concentrations is detectable, although it will be degenerate with other exoplanet properties. A potentially important effect of photochemistry on transmission spectroscopy, which is, however, not probed by our chemical network is photo-ionization and the increased abundance of H$^-$. The latter has been shown to be an important opacity source in the hydrogen-rich atmospheres of exoplanets \citep{Arcangeli2018, Lewis2020, Rathcke2021}.

In high-resolution spectroscopy observations, which typically probe higher altitudes than transmission spectroscopy, photochemistry potentially has a bigger role to play. \chem{HCN} and \chem{C_2H_2} have already been detected in some exoplanets \citep{Hawker2018, Cabot2019, Giacobbe2021}, and are being used to constrain their elemental abundances. Photochemistry, and disequilibrium chemistry in general, are important to understand such results. In addition, a pseudo-2D or 3D framework is particularly favourable to interpret the spatial distribution of chemical species at high altitudes, which can also be probed with high-resolution spectroscopy \citep{Ehrenreich2020, Kesseli2021}.

Finally, we note that in our models, we have adopted constant stellar spectra and discussed time-invariant, steady-state chemistry. As such we did not incorporate the effect of stellar flares, which could alter the chemical composition of exoplanet atmospheres on short to long timescales \citep{Venot2016}.
We will investigate the impact of stellar flares on exoplanetary chemistry in a 2D context in an upcoming study (Konings et al., \emph{subm.}).

\section*{Acknowledgements}

We would like to thank Paul Brandon Rimmer, Antonio Garc\'ia Mu\~{n}oz, J\'er\^{o}me Loreau, Conny Aerts, Jon Sundqvist, and Minh Tho Nguyen for insightful comments and discussions. We further thank Marcelino Ag\'undez for helpful advice regarding the pseudo-2D code, and the anonymous referee for useful suggestions.
RB acknowledges funding from a PhD~fellowship of the Research Foundation -- Flanders (FWO).
OV acknowledges support from the Agence Nationale de la Recherche (ANR), through the project EXACT (ANR-21-CE49-0008-0) and from the CNRS/INSU Programme National de Plan\'etologie (PNP). This work was supported by CNES, focused on the EXACT project and Ariel.
LC acknowledges support from the DFG Priority Programme SP1833 Grant CA 1795/3 and the UK Royal Society Grant URF R1 211718.
LD acknowledges support from the FWO research grant G086217N.
RB, TK, and LD acknowledge support from the KU Leuven IDN/19/028 grant ESCHER.

%%%%%%%%%%%%%%%%%%%%%%%%%%%%%%%%%%%%%%%%%%%%%%%%%%%%%%%%%%%%%%%%%%%%%%%%%%%

\section*{Data Availability Statement}

The data underlying this article will be shared on reasonable request to the corresponding author.

%%%%%%%%%%%%%%%%%%%%%%%%%%%%%%%%%%%%%%%%%%%%%%%%%%



\bibliographystyle{mnras}
\bibliography{photochem} % if your bibtex file is called example.bib

% \begin{thebibliography}{99}
% \bibitem[\protect\citeauthoryear{Author}{2012}]{Author2012}
% Author A.~N., 2013, Journal of Improbable Astronomy, 1, 1
% \bibitem[\protect\citeauthoryear{Others}{2013}]{Others2013}
% Others S., 2012, Journal of Interesting Stuff, 17, 198
% \end{thebibliography}

%%%%%%%%%%%%%%%%%%%%%%%%%%%%%%%%%%%%%%%%%%%%%%%%%%

%%%%%%%%%%%%%%%%% APPENDICES %%%%%%%%%%%%%%%%%%%%%

% \appendix

% \section{Some extra material}

% If you want to present additional material which would interrupt the flow of the main paper,
% it can be placed in an Appendix which appears after the list of references.

%%%%%%%%%%%%%%%%%%%%%%%%%%%%%%%%%%%%%%%%%%%%%%%%%%

% Don't change these lines
\bsp	% typesetting comment
\label{lastpage}
\end{document}